# Ballistic magnetoresistance in nickel single-atom conductors

Matthew R. Sullivan, Douglas A. Boehm, Daniel A. Ateya, Susan Z. Hua, and Harsh Deep Chopra[*]

*Thin Films and Nanosynthesis Laboratory, Materials Program, Mechanical & Aerospace Engineering Department, SUNY-Buffalo, Buffalo, NY 14260*

*Abstract*— Large ballistic magnetoresistance (BMR) has been measured in Ni single-atom conductors electrodeposited between microfabricated thin films. These measurements irrefutably eliminate any magnetostriction related artifacts in the BMR effect. By making measurements on single atom conductors, the *benchmark* for the incontrovertible evidence against magnetostriction is set at the unyielding condition of the known quantum mechanical principles, namely, $1\,G_o = 2e^2/h = 1/12900\,\Omega^{-1}$ (or $\frac{1}{2}G_o = 1/25800\,\Omega^{-1}$ in a spin-split state) is the universal threshold ballistic conductance of an *unbroken* single atom contact below which even an angstrom separation of the contact due to magnetostriction is immediately signaled by an abrupt and large increase in tunneling resistance of several hundred thousand ohms across the gap. The present approach to electrodeposited point contacts between microfabricated thin films also provides an independent confirmation of Garcia's original BMR experiments on atomic point contacts that were made by a mechanical method [N. Garcia et *al*. Phys. Rev. Lett. **82**, 2923 (1999)].

Spin dependent electron transport across a single atom represents the ultimate miniaturization of spintronics devices. Garcia *et al.*[1] reported several hundred percent ballistic magnetoresistance (BMR) in atomic point contacts of Ni made by a mechanical method, and subsequently observed 400-700% BMR in electrodeposited Ni nanocontacts made between bulk Ni lead wires.[2,3] Large magnetoresistance effect has also been reported in half-metal magnetite ($Fe_3O_4$) point contacts made by a mechanical method.[4] Using experiments similar to Garcia's, we have recently reported several thousand percent BMR effect in electrodeposited Ni nanocontacts formed between bulk Ni wires.[5,6] Similar BMR values have since been reported.[7]

While much progress has been made in recent years towards understanding the origin of the BMR effect, it is also accompanied by uncertainty that invariably follows the maturation of any new field or discovery. For BMR, a principal difficulty lies in separating the electronic origin of the effect from signals that could possibly be attributed to magnetostriction. In this regard, study of nanocontacts deposited between substrate-constrained thin films is an important first step in removing this uncertainty. However, it is not sufficient in itself. Experiments must be designed in which the contraction or elongation of the nanocontact due to magnetostriction would immediately and unambiguously manifest itself; conversely, its absence would prove the electronic origin of the BMR effect.

Fortunately, such an unambiguous experiment can be designed by exploiting the quantized nature of conductance in ballistic conductors made of a single atom.[8] In metals where the Fermi wavelength of the electrons is typically of the order of 0.5 nm, the transverse confinement of the electron wave functions by the narrow contact diameter quantizes the energy levels (channels), with each channel contributing to the total conductance in discrete units of $2e^2/h$ ($\equiv G_o$); $e$ is the charge of the electron and $h$ is the Planck's constant.[9-11] In the case of a ferromagnetic contact, the spin degeneracy can be lifted, causing the stepwise change in conductance to occur in units of $\frac{1}{2}G_o$ instead of $G_o$;[12,13] Spin-splitting and magneto-conductance has even been reported in (oxidized) Cu point contacts.[14,15] The separation between the channels increases with a decrease in the contact diameter, causing an increasing number of channels to rise above the Fermi level and become unavailable for conductance. In the limit of a single atom contact, only a single channel is left below the Fermi level, for a total conductance of $1\,G_o$ $(=2e^2/h=1/12900\,\Omega^{-1})$ - the resistance of a single (monovalent) atom, see Refs. [16-18]. In other words, the conductance plateau of $1\,G_o$ =1/12900 $\Omega^{-1}$ (or $\frac{1}{2}G_o$ =1/25800 $\Omega^{-1}$ in a spin-split ferromagnetic contact) is a *universal* and *discrete threshold* for an unbroken atomic point contact. As soon as the contact is broken, the ballistic transport with its transversally quantized wave functions is replaced by tunneling where the wave functions are now longitudinally confined across the narrow gap. This transition is signaled by an abrupt and large exponential increase in resistance from the universal ballistic threshold of 12900 $\Omega$ (or 25800 $\Omega$) to tunneling resistance of several hundred thousand ohms. *This is an unyielding condition of the known quantum mechanical principles, and it unambiguously determines whether the atomic point contact is broken or intact*. Nickel has negative magnetostriction and it contracts in the direction of an applied magnetic field. Therefore even an angstrom separation of the single atom Ni contact from adjacent Ni leads due to magnetostriction will immediately signal a broken contact. Hence magneto-conductance measurements on single atom Ni contacts offer an unambiguous experiment in ascertaining the absence or presence of magnetostriction related artifacts in BMR (magneto-conductance, the inverse of magnetoresistance is measured, since conductance rather than resistance is traditionally measured in these quantum conductors). The results of these experiments on single atom Ni contacts deposited between Ni thin film leads in the present study confirm the electronic origin of the BMR effect.

Single atom Ni contacts were electrodeposited between microfabricated thin-film leads of Ni deposited on silicon wafers. High resistance <100> silicon wafers coated with an additional 1 μm thick silicon dioxide insulator layer were used to suppress any possible leakage currents through the substrate during subsequent electrodeposition and testing. Special microfabrication steps were taken to keep the starting gap between the patterned thin film leads to be as small as possible (1 μm) in order to ensure that the leads are maximally constrained by the substrate. This was achieved by using photolithography to first form 0.4-0.5 μm deep channels on the silicon wafer by reactive ion etching, as shown in the scanning electron microscope (SEM) image in Fig. 1(a), followed by deposition of a Ta (3 nm)/Ni (10 nm) seed layer within the channels by dc magnetron sputtering. The Ni seed layer was then thickened by electrodeposition until a sub-micron gap was left between the leads, as shown in the SEM micrograph in Fig. 1(b). Since the 0.4-0.5 μm high channel walls initially restrict the lateral growth of the films, sufficiently thick Ni films (2-3 μm) with low lead resistance (3-5 Ω) could thus be electrodeposited. Also note the patterned tip in Fig. 1(a) to facilitate the subsequent formation of the atomic contacts across it. The thickened Ni film in Fig. 1(b) retains this patterned tip geometry (except for the inevitable rounding) illustrating the almost layer-by-layer electrodeposition growth conditions. The final contact is made by thickening the lead with the tip as cathode and the opposite lead as anode, as shown in the SEM micrograph in Fig. 1(c). Nickel was electrodeposited at room temperature using nickel sulfamate solution (84 g/l Ni as metal in $Ni(SO_3NH_2)_2.2H_2O$; 30 g/l boric acid, pH 3.3) at a deposition voltage between -0.7 and -1.0 V versus a saturated calomel electrode. The single atom conductors were made using the self-terminating method of Tao.[19] In particular this simple and elegant method allows electrodeposition of stable nanocontacts to a size-resolution of a single atom. Following the formation of the contact the electrical measurements were made at a voltage ranging from 200 mV to 350 mV.

Figure 2(a) shows an example of a stable single atom Ni conductor at a conductance plateau of $1\,G_o$ made by the self-terminating method. Figure 2(b) shows another example of a stable single atom Ni conductor that was grown to the targeted conductance plateau of $1\,G_o$. In Fig. 2(b) the conductance curve during the growth of the contact rises to the stable $1\,G_o$ plateau in a stepwise fashion with the brief appearance of the $\frac{1}{2}\,G_o$ plateau even though no magnetic field was present during the electrodeposition. Previously, half-integral multiples of $G_o$ have been observed in Ni nanocontacts formed (by mechanical methods) in the presence of an applied magnetic field, and attributed to a field-induced single domain state in the Ni wires adjacent to the contact.[12,13] In our study it was found that even in the absence of an applied magnetic field the conductance often stabilized spontaneously at half-integral multiples of $G_o$.[20] Therefore the spontaneous appearance of such spin-split states in contacts made in zero-field is indicative of a single domain state in the vicinity of the contact, and is not entirely unexpected given the overall small dimensions of the contact and the thin film leads in the vicinity of the contact. For example, Fig. 2(c) shows a conductance plot for a single atom contact that has spontaneously stabilized in a spin-split state of $\frac{1}{2}\,G_o$. As shown in Fig. 2(c), initially when a gap is present between the leads, the resistance is very high (0.5-1.0 M$\Omega$). As soon as the contact is formed, the conductance makes an abrupt jump and spontaneously stabilizes at the $\frac{1}{2}\,G_o$ conductance plateau. Figure 2(d) shows another example of a single atom contact that has stabilized in its $\frac{1}{2}\,G_o$ spin-split state.

Figures 3(a-b) show respectively the magneto-conductance of single atom Ni conductors stabilized at $\frac{1}{2}G_o$ and $1\,G_o$ conductance plateaus. In addition, Figs. 3(c-d) show the magneto-conductance of samples that were stabilized at $\frac{3}{2}G_o$, and $2\,G_o$ plateaus, respectively. Each sample in Figs. 3(a-d) was stabilized at its respective conductance plateau for a sufficiently long period of time prior to applying the magnetic field. The direction of the applied field is along the axis of the contact joining the thin film leads, as shown in the SEM micrograph in Fig. 1(c). As shown in Fig. 3(a), the conductance of the single atom conductor changes from a stable value of $\frac{1}{2}G_o$ in zero field to $5\,G_o$ at 500 Oe. Once the field is removed the conductance drops back to its zero-field plateau. Since the sample has low conductance (or high resistance) at zero-field and a high conductance (lower resistance) at high field, this represents a negative BMR effect of 900%.

Figure 3(b) shows another single atom conductor that was stabilized at $1\,G_o$ prior to applying the magnetic field. As in Fig. 3(a) the sample in Fig. 3(b) also shows an increase in conductance with increasing field, rising from $1\,G_o$ at zero-field to $3\,G_o$ at 500 Oe, and dropping back reversibly to $1\,G_o$ once the field is turned off, giving a 200% BMR effect. The samples stabilized at $\frac{3}{2}G_o$, and $2\,G_o$ plateaus in Figs. 3(c-d), respectively, similarly show a negative BMR effect of 200% and 2000%, respectively. Note that the conductance plateaus of $\frac{3}{2}G_o$, and $2\,G_o$ also likely represent the magneto-conductance behavior of single Ni atom conductors albeit in different electronic different, see Refs. [16-18], since the open $d$ shell of Ni is capable of carrying at least five channels/atom (even without considering the effect of $s$ and $p$ orbitals, which can potentially lead to an even greater number of channels per Ni atom). Also, Fig. 3(c) is shown deliberately to underscore the value of making magneto-conductance measurements at $\frac{1}{2}G_o$ and $1\,G_o$ as in Figs.

3(a-b), since the dip in conductance in Fig. 3(c) preceding the eventual increase in conductance would be otherwise be harder to interpret due to magnetostriction related artifacts.

Nickel has negative magnetostriction (saturation magnetostriction strain $\lambda_s = -34 \times 10^{-6}$).[21,22] Therefore a first step in minimizing the role of magnetostriction is to prevent the sample from freely changing its physical dimensions. In the present study, the microfabricated thin film leads are constrained by the substrate and by the microfabricated channels in which the leads are partially embedded. This constraint is further maximized by minimizing the starting gap between the leads, which is 1 μm as shown in Fig. 1(a). Only in this short segment the leads have no seed layer underneath. The maximum contraction strain for every μm of free length $l$ is given by $\Delta l = \lambda_s l$, giving a maximum possible elongation of -0.034 nm, or ~ -0.3 Å, which is only a fraction of the atomic diameter of Ni.[22] Even the presence of this minuscule elongation cannot be ruled out *a priori*, and hence the necessity of using single atom conductors in making an unambiguous interpretation. Any residual magnetostriction contraction in the leads due to this micron long segment would tend to break the contact and lower the conductance. Since the measurements were made on single atom conductors, this would be signaled by a large and instantaneous drop in conductance of the order of several hundred thousand ohms even for an angstrom separation. *Contrary to this*, the magneto-conductance of single atom conductors in Fig. 3 *increases* rather than decreases in response to an applied magnetic field, conclusively ruling out the role of magnetostriction in these measurements. The results also show an impressive mechanical robustness of these seemingly fragile entities. Considering that in terms of size these single atom Ni conductors are most delicate compared to their larger counterparts, they also represent the litmus test for the susceptibility of larger contacts to break apart due to magnetostriction.

It is also important to note that the first reported BMR measurements by Garcia were done on atomic point contacts of Ni between bulk Ni wires using a mechanical method.[1] The present study uses an entirely different experimental technique to reproduce Garcia's results, an important yardstick to resolve any scientific controversy.

Finally, in our previous papers on Ni nanocontacts made by electrodeposition between constrained bulk wires.[5,6] we had attributed the behavior of larger contacts as arising from a multiplicity of atomic point contacts due to the presence of microscopically sharp asperities at the tip of the wires, and giving an *apparent* net resistance (1-100 $\Omega$) equivalent to that of a monolithic classical conductor.[6] This is the source of some confusion and the use of bulk contacts of equivalent resistance is not a substitute for true ballistic conductors. Within a contact, the presence of a magnetic dead layer enhancing spin polarization and producing large BMR values has been proposed.[23] Experiments are currently underway to understand the singular and collective behavior of these contacts.

In conclusion, the present study on single atom Ni conductors irrefutably rules out any magnetostriction related artifacts in the measured BMR effect.

This work was supported by NSF-DMR-0305242 and NSF-DMR-97-31-733, and this support is gratefully acknowledged. Microfabrication was done at the Cornell Nanofabrication Facility (a member of the National Nanofabrication Users Network), which is supported by the National Science Foundation Grant ECS-9731293 and Cornell University.

[22]This of course is the upper limit since attaining saturation magnetostriction strain for Ni requires upwards of ~3000 Oe (based on our magnetization measurements), whereas only 1/6[th] of this value (500 Oe) was applied in the present study. In the presence of mechanical constraints preventing the Ni from changing its physical dimensions (such as the constraints imposed on thin film leads by the underlying substrate), the unrealized magnetostriction strain manifests itself as equivalent stress $\sigma^H = E\lambda^H$, where $E$ is the Young's modulus (=20 GPa for Ni) and $\lambda^H (\leq \lambda_s)$ is the magnetostriction at field $H$.

[23]N. García, G. G. Qiang, and I. G. Saveliev, Appl. Phys. Lett. **80**, 1785 (2002).

# FIGURE CAPTIONS

**FIG. 1.**    (a) SEM micrograph showing 0.4-0.5 μm deep microfabricated channels on oxidized silicon wafer in which Ni thin film leads were subsequently electrodeposited. Note the 1 μm initial gap between the channels. (b) SEM micrograph showing thin film leads made by electrodeposition of Ni on a previously sputter deposited Ta/Ni seed layer inside the channels. (c) The final contact between thin film leads is made by electrodeposition of Ni until the lead with the tip joins the opposite lead.

**FIG. 2.**    Conductance plots for four different single atom conductors stabilized at (a-b) $1\,G_o$ and (c-d) $\frac{1}{2}\,G_o$ conductance plateaus.

**FIG. 3.**    Magnetoconductance of single atom Ni conductors stabilized at (a) $\frac{1}{2}\,G_o$, (b) $1\,G_o$, (c) $\frac{3}{2}\,G_o$, and (d) $2\,G_o$.

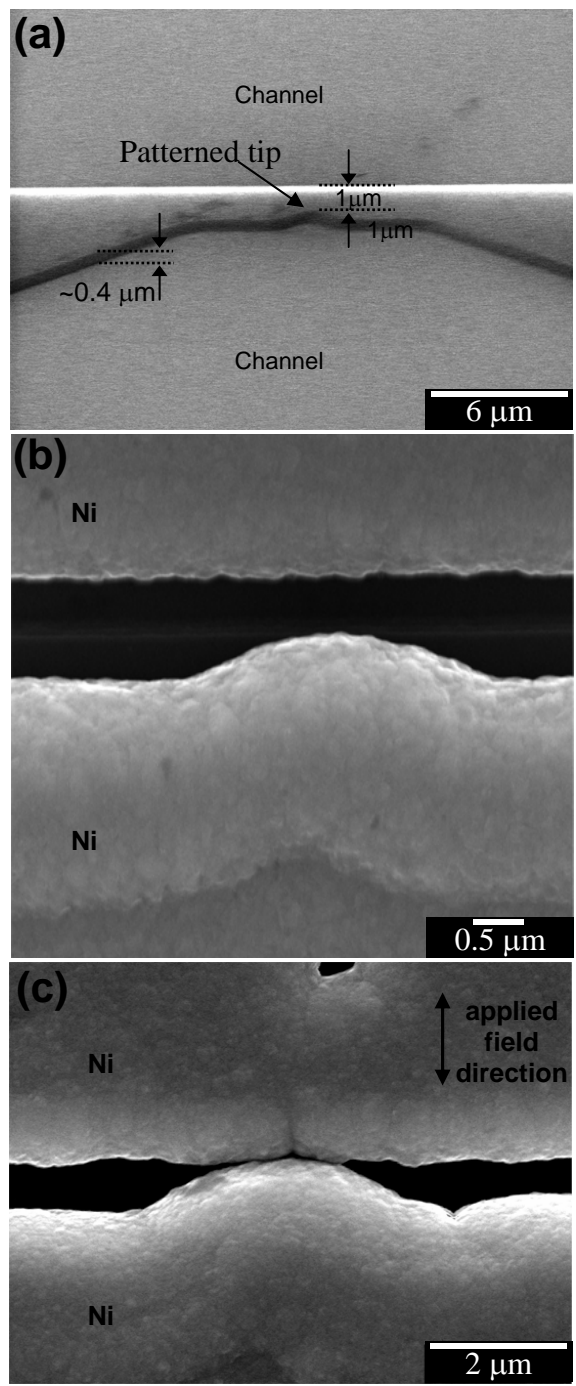

**Figure 1**

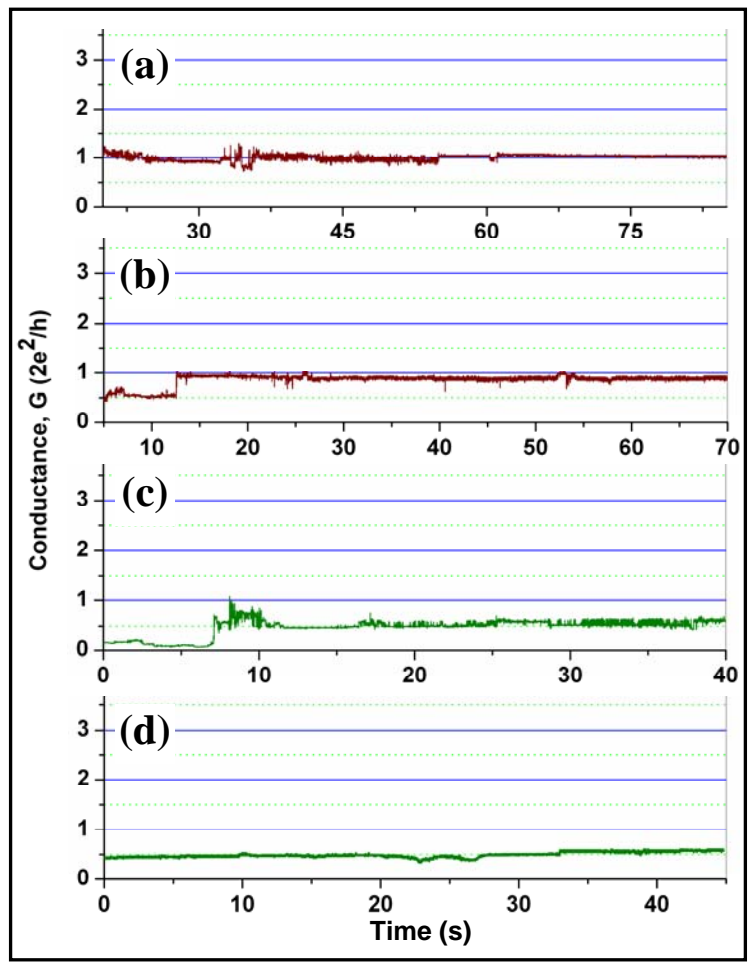

**Figure 2**

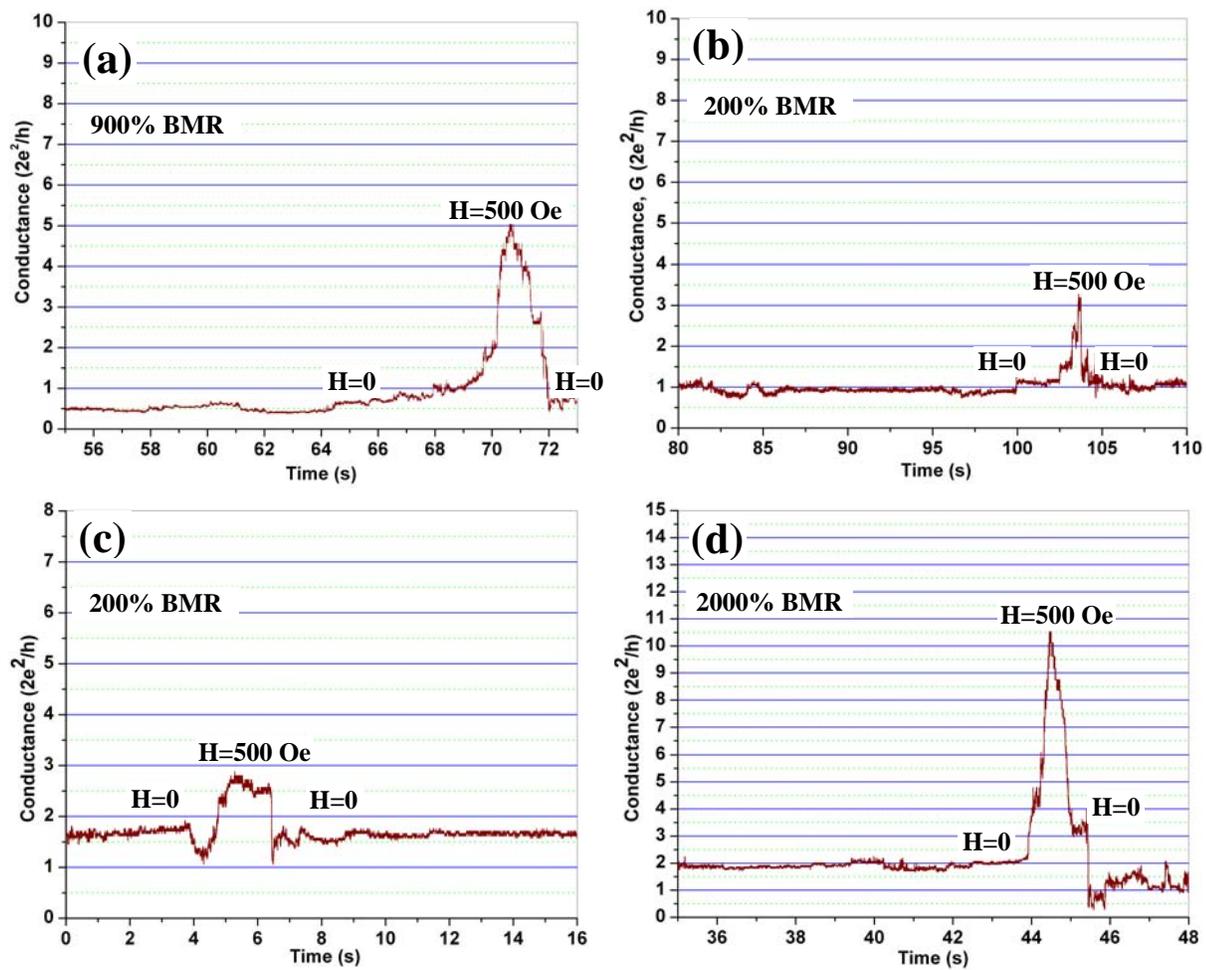

**Figure 3**